# DESIGN EVALUATION OF SOME NIGERIAN UNIVERSITY PORTALS: A PROGRAMMERS POINT OF VIEW

[1]Abdulhamid, Shafi'i Muhammad, [2]Ismaila, Idris
[1,2]Dept. of Cyber Security Science, Fed. Univ. of Techno. Minna, Niger State, Nigeria.
[1]shafzon@yahoo.com, [2]ismi_idris@yahoo.co.uk

*Abstract*

*Today, Nigerian Universities feel pressured to get a portal up and running – dynamic, individualized web systems have become essential for institutions of higher learning. As a result, most of the Nigerian University portals nowadays do not meet up to standard. In this paper, ten (10) Nigerian University portals were selected and their design evaluated in accordance with the international best practices. The result was revealing!*

**Keywords:** *Portal, Database, Website, Personalization, Internet and System design.*

### Introduction

The term "portal" is so overused as to be meaningless. It has gotten so bad that many places build a portal simply by putting the word "portal" on their normal web pages. We need a better definition.

As defined by IBM, an Internet portal is "a single integrated, ubiquitous, and useful access to information (data), applications and people." A portal may look like a Web site, but it is much more. A portal is a gate, a door, or entrance. In the context of the World Wide Web, it is the next logical step in the evolution toward a digital culture. Portals are a way of bringing together all the information that users need in a single place, accessible in a coherent way that provides for enhanced productivity. It is an application that has link with a database. It is not like a static website we know, it is dynamic in the sense that a portal is more interactive whereby we can actually input data and also get the information we are looking for in it. In other words, a website is a gateway to a portal, meaning that we see a portal in a website.

A University Portal or Institution Portal is a one-stop client-oriented web site that personalizes the portal's tools and information to the specific needs and characteristics of the person visiting the site, using information from the faculty databases. In a normal University web site, the information is out there, but it is up to the visitor to find it. They have to either use a search engine, or navigate the links on the University's web page [Cronin, 2001]. Often people miss information they couldn't find. In a "University Portal", the visitor identifies himself to the portal. The portal then uses the detailed knowledge the faculty has about this person to gather together all the information relevant to that person and display it in one place. This information could be generic publicly accessible information, or it could be confidential information specific to that individual (like course grades or payroll information). The emphasis has shifted from a public web site showing on-line pamphlets to a user-oriented web site that provides tools, reports, and services specifically designed for that individual.

### Benefits of Portals to the University

The potential benefits realized from a campus-wide portal are extensive and revolve around increased efficiency and the ability to provide university constituencies with access to information. Some notable benefits include:
- User identification through a single log-on account;

- Centralized administration and enforcement of security driven by user authentication protocols;
- A common entry point for all services available at the university;
- A common architecture for integrating administrative systems, course management tools, and content from disparate systems and providers;
- Consistent appearance and behaviour, facilitating improved user experience and improved relationship with and between constituents; and
- The ability to engage, empowers, and retains constituents by providing a sense of membership in an academic community.

A successful portal implementation requires a significant amount of effort and collaboration between key stakeholders at the university. Without the full commitment from senior university management and other key stakeholders, the project may never return its full benefit.

**Literature Review**

[Daigle and Cuocco, 2002] believe that a major reason for deploying portals is "to improve productivity by increasing the speed and customizing the content of information provided to internal and external constituencies." They also suggest that portals serve a knowledge management function by "dealing with information glut in an organized fashion."

Web portals have been used to streamline and automate administrative functions in higher education. The most recent application of portals in higher education has been to create a point of access for administrative functions for students, such as registration, financial aid and academic records, or for staff, such as timesheets, leave balances and the like [Olsen, 2002]. In this way, use of portals maximizes efficient use of staff and students' time [Pickett, 2002].

A university portal potentially offers other stakeholders a vital link into the university. Parents are eager to see what their children are experiencing. Citizens and state legislators are very interested in what their tax dollars are being used for and how the university can contribute to the state's well-being and economic improvement. [Katz et al, 2000] contend, "The new, wonderful, and challenging aspect of Web management posed by portals is the idea of creating and managing information systems whose primary purpose is to sustain positive relationships between an institution's stakeholders and the institution. That's new." They further suggest that portals represent new strategic means of increasing a university's competitive position by fostering innovation and research activities that can lead to greater acquisition of grants and improved prestige for the university. By harnessing the ability of portals to create learning and research communities, portals can further leverage the huge intellectual capital based contained within the organization via collaborative, synergistic activities.

Portals also serve to empower individuals within a more broadly defined university community. By providing easy accessibility to both explicit and tacit knowledge as well as communities of practice, people are not constrained by geographic or other physical barriers in terms of communicating and exploring new knowledge. "The portal will improve the efficiency of knowledge exchange and deliver a set of shared business objectives that include communications around best practices, a gateway to research on the use of teaching and learning through technology, professional development, policy development and review and resource development" [Kidwell, 2000]. Portals facilitate knowledge transfer through the inclusion of multiple communication channels, such as message boards and directories; moving beyond the one-sided information exchange found in traditional web sites.

**System Design Life Cycle**

The design of the system was based on fundamental system analysis and design principles (Figure 1) with some deviations to accommodate the unique aspects of the project. Table 1 shows a summary of activities for this work in the development of the university portal framework.

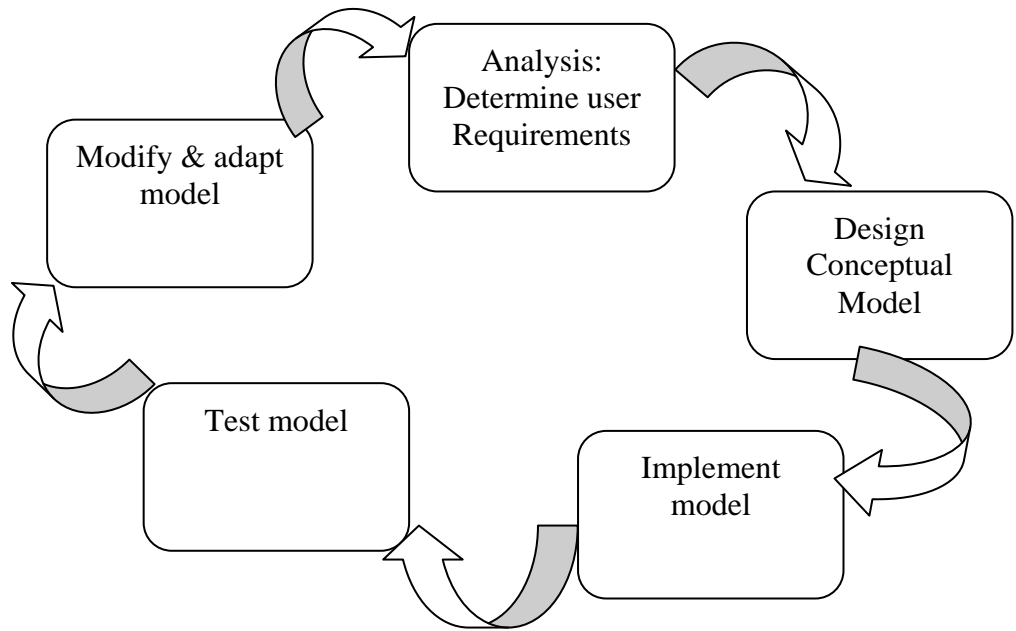

**Figure 1**: Systems design life cycle

| Task | Description |
|---|---|
| Benchmark against research portals of excellence | Surfed the Web to find research portals throughout the world and saved those that were excellent in terms of organization, content, and navigability. |
| Marketing research | Interviewed faculty members from different disciplines across the university to determine their needs / user requirements for the portal |
| Develop conceptual model of portal | • Developed an entity relationship diagram to show how the parts of the portal are interrelated.<br>• Developed a flowchart to reflect the user requirements and organization of the site. |
| Build portal prototype | • Developed a storyboard for site development.<br>• Downloaded the content management system (CMS) to use as functional shell for portal<br>• Customized CMS to reflect needs of users. |
| Portal demonstration | Presented portal prototype to top management for organizational buy-in. |

**Table 1**: Summary of project activities

**Evaluation Criteria (Best Practices)**

In an effort to best understand our universities existing systems and processes, and to avoid redundancies in future systems and processes, everyone whose job impacted the admission enrolment process need to be invited to attend the needs assessment. The following are best practices that were used to guide the discussions and to customize the various features within the university portal.

### a) Personalize

University portals are designed to make the member's experience with the portal as personalized as possible. Based primarily on information collected in the sign-in form, the system pushes information specific to a prospect's background and interests. This makes the system easier to use and provides a high level of customer service that is likely to influence the prospect's decision to apply or register.

Via sign-in, students have access to the following personalized features:
- Members greeted by name
- Personalized content (text)
- Username and password retrieval
- Online admissions application with personal details pre-populated in fields
- Automated FAQ's
- Campus tour registration with personal details pre-populated in fields
- Package request with personal details pre-populated in fields
- E-mail account
- Mailbox (full history of correspondence within Connect2Concordia.ca)

### b) Timely and Relevant

As the higher education landscape becomes more and more competitive, technologies develop that make prospective students more discerning than ever. Many students use e-mail and the Internet as a primary method of contact. This being the case, it is critical that we give prospects the information they are looking for, as quickly as possible. This is achieved through the following system features:

- Prospective students' button: We included a button on our home page for prospective students to bring them to the portal. For brand consistency, the portal has the same look and feel as our Web site.
- Automated e-mail responses: Ensure that students and staff will receive an immediate response to every question or action. For example, if a prospect requests a package, the system will thank them for that request immediately, and will let them know when they can expect to receive the package in the mail.
- Refresh pages: This pertains to the sign-in. When a prospect indicates they are from the Niger State, for example, the screen will refresh to load in a drop down of Niger States Local Governments. In this way, the user is only required to see fields that are applicable to them.
- Automated FAQ's: Give prospects immediate, on-screen responses to their questions.
- Content based on segments: The system is designed to push content relevant to a prospect's background and interests. For example, international prospects will see different admissions information than a domestic prospect.
- On-line Campus Tour registration: As prospects expect immediate service, they should not be required to call or send an e-mail to register for an event. This section of the site allows them to register for, re-schedule or cancel Campus Tours.
- Public information: General Information related to Programs, Admissions, Financial Aid, and Accommodation are available to prospective students without requiring sign-in (a prospect may need to know if we offer a specific program, for example, before they consider making an inquiry).
- Dropdowns: This limits the amount of typing required by a prospective student. It also ensures a clean database on the back-end.

- Pre-populated fields: Fields are pre-populated as often as possible, so prospects have to enter their personal information only once.
- Smart navigation: Higher Education portals, main sections, such as events, program information, financial aid, accommodation, etc., are clearly displayed and are part of the standard navigation on every page. Sections that require sign-in are grouped together and marked with a small lock.

### c) Building Relationships Using E-mail, Mail and Telephone

The Higher Education portals supports the idea that regular, consistent, and targeted communication with both students and staff builds relationships and ultimately, encourages enrolments. The system promotes communication via e-mail, letter, or phone call.

Targeted communication: This tool allows us to create targeted communication campaigns to send to members. After creating a specific group, using filters based on characteristics of the member (i.e. program interest, prospect type, location), a campaign can be deployed.

### d) Track Inquiries

The Higher Education portals are designed to track communication with students by creating a student record for each. Members create the record for themselves via the portal, or a portal administrator creates the record for the student or staff when they receive a phone call, fax, walk-in, etc. The following features / efforts support inquiry tracking:

- One point of entry: Strive to funnel as many members to the portal as possible. This process began, by reviewing the Higher Education website (if any) and redirecting links to the portal. Inquiries from outside sources such as school directories are directed to the portal as well.
- Prospect record lookup / Creation: This section allows us to create or view member's record. In order to view a member record, the portal administrator uses a look-up feature.
- How did you hear about us: This question is asked (via dropdown) of every member that signs into our system. This can easily add to a dropdown of choices to keep it up to date with marketing and recruitment efforts.
- Communication history: The portal administrator can view the communication history with each student or staff. This includes incoming messages from the member portal and outgoing e-mail and letters, the questions they asked, the answers they received, call notes, and campus tours attended.
- School/Institution Records: This section allows portal to create or view a school record. Each record contains address information, school visit details (type of visit and visit records), notes and multiple contacts at the school. Each contact record includes name, title, e-mail address, etc.

### e) Integration

The online application for admission should also be integrated into the Higher Education portals. This will allow the school to make it that much easier to convert an inquiry to an applicant. It is also good to integrate the school records.
- Reporting: The Reports section in the Administrator Site allows Higher Education portals staff to view trends and statistics based on the profile of the students who registered through the portal. This includes reports on geographical distribution, ad hoc reports, measuring free-type e-mails, FAQ usage, and campus tour registration and participation.

### f) Make the System Easy to Manage

We wanted a system that will be easy to manage.

- Content management: Staff can manage the content on Higher Education portals without the need for technical skills. Using the system's content management tools in the Data section of the Administrator site, it should be easy to add, edit, hide or delete pages.
- Templates: Templates are used to create the e-mails and letters that are attached to package requests, inquiries to answer, and targeted e-mail and letter campaigns. They may be used to manage individual member records or multiple member records all at once.
- Combined databases: The system combines all of our student databases into one. This streamlines all processes and avoids duplication of efforts.

**Assessing Some Nigerian University portals**

In this section, ten (10) Nigerian University portals are selected for design assessment which is based on the best practices mentioned above. The result is presented in tabular form and then a chart drawn for the data presented in the table 2.

| | Nigerian Universities Portals | 1. F.U.T. Minna, www.futminna.edu.ng | 2. B.U.K. Kano, www.bukportal.com | 3. A.B.U. Zaria, www.abuportal.net | 4. University of Nigeria, Nsukka, http://portal.unn.edu.ng | 5. University of Abuja, www.unibujaportal.com | 6. Madonna University Elele, Rivers, www.madonnauniversityportal.com | 7. University of Lagos, http://unilagcomm.com | 8. University of Ibadan, http://www.ui.edu.ng | 9. University of Benin, http://www.uniben.edu | 10. Ladoke Akintola University of Technology, Ogbomoso. http://www.lautech.edu.ng/ |
|---|---|---|---|---|---|---|---|---|---|---|---|
| | **Best Practices** | | | | | | | | | | |
| A | Personalization (10) | 7 | 7 | 7 | 7 | 8 | 6 | 8 | 8 | 6 | 7 |
| B | Timely and Relevant(10) | 4 | 4 | 4 | 3 | 4 | 2 | 4 | 3 | 3 | 3 |
| C | Building Relationships Using E-mail and Telephone (10) | 5 | 3 | 5 | 3 | 3 | 4 | 5 | 6 | 5 | 4 |
| D | Track Inquiries (10) | 6 | 5 | 3 | 5 | 2 | 7 | 4 | 4 | 3 | 5 |
| E | Integration (10) | 5 | 5 | 4 | 6 | 4 | 6 | 5 | 4 | 5 | 4 |
| F | Make the System Easy to Manage (10) | 6 | 7 | 6 | 6 | 6 | 6 | 7 | 7 | 8 | 7 |
| | **Total Scores(60)** | 33 | 31 | 29 | 30 | 27 | 31 | 33 | 32 | 30 | 30 |

**Table 2:** Some Nigerian University Portals – Assessment Table
All data taken between 01/03/2009 and 01/07/2009.

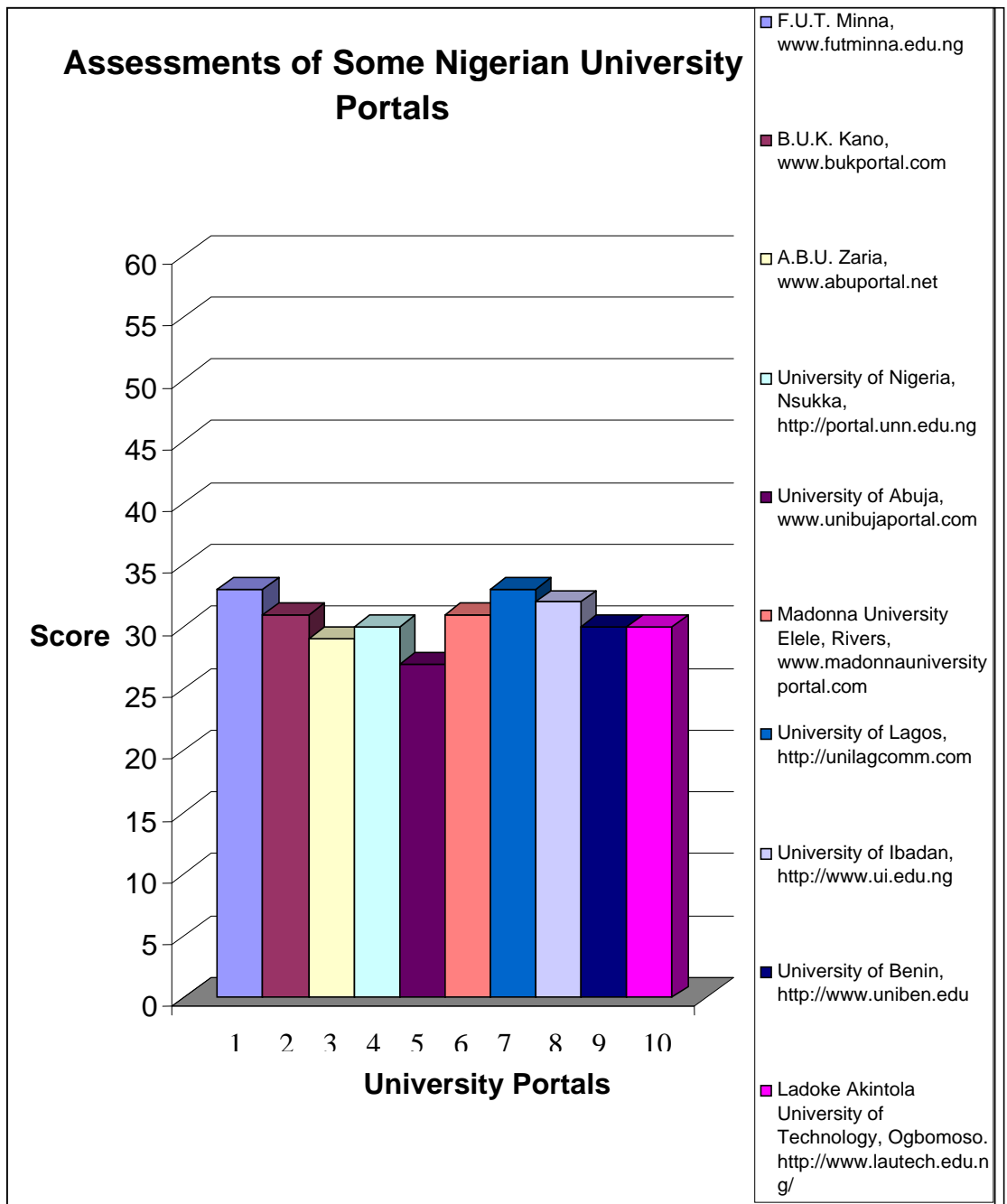

**Figure 2:** Assessment of Some Nigerian University Portals Chart

**Discussion of Results**

Generally, from the analysis of the data in the table2 and figure2 above, it is easy to say that most Nigerian University portals are averagely designed. Thus, they don't take full advantage of the functions which a well designed portal can provide. It is also easy to deduct that the University of Lagos and Federal University of Technology Minna portals have one of the best portal design within the country. The University of Abuja portal design comes under our hammer for relatively poor design.

**Conclusion and Recommendations**

In this internet driven age, Nigerian Universities need to come out and make a fashion statement by designing some world class portals that can compete with any other anywhere in the world, in accordance with the international best practices. As the saying goes "anything that is worth doing, worth doing well", so designing an average portal to serve the basic need of registration the universities is not the only function which a portal can perform.

Therefore, we recommend that Nigerian Universities should design their portals in line with international best practices and avoid unethical designs. The secret behind any good application software is good design. Without a good design, coding and maintenance becomes a tedious task.

Lastly, we must also commend some of the portals for making provisions for increments modules to be added with time. By doing so, the portals evolve with time and become more enriched and more standardized.

**References**

1. Cronin, B. (2001) 'Knowledge management, organizational culture and Anglo-American higher Education', *Journal of Information Science*, Vol. 27, No.3, pp.129-137.
2. Daigle, S.L. and Cuocco, P.M. (2002) 'Portal Technology Opportunities, Obstacles, and Options: A View from the California State University'. In *Web Portals and Higher Education, Technologies to Make IT Personal* (109-123). Boulder, CO: Educause. Retrieved 10, January 2005 http://www.educause.edu/ir/library/html/pub5006.asp
3. Olsen, F. (2002) 'The Power of Portals'. *Chronicle of Higher Education*, 48(48), A32-A34. Retrieved via *Academic Search Premier* University of Maine 27 January 2005 <www.library.umaine.edu>.
4. Pickett, R.A. and Hamre, W.B. (2002). Building Portals for Higher Education. *New Directions for Institutional Research*, Vol. 113, pp.37-55.
5. Katz, R.N. and Associates. (2002) 'Web Portals and Higher Education, Technologies to Make IT Personal'. Boulder, CO: Educause. Retrieved 10 January 2005 <http://www.educause.edu/ir/library/html/pub5006.asp>
6. Katz, S.N. (2002) 'The Pathbreaking, Fractionalized, Uncertain World of Knowledge'. *Chronicle of Higher Education*, Vol. 49, No. 4, pp.B7-B10.
7. Kidwell, J.J., Vander Linde, K.M. and Johnson, S.L. (2000) 'Applying Corporate Knowledge Management Practices in Higher Education'. *Educause Quarterly*, Vol. 4, pp.28-33.